\documentclass[showpacs,showkeys,preprintnumbers,amsmath,amssymb]{revtex4}

\newcommand{\sect}[1]{\setcounter{equation}{0}\section{#1}}
\renewcommand{\theequation}{\arabic{section}.\arabic{equation}}
\newcommand{\subsect}[1]{\setcounter{equation}{0}\subsection{#1}}
\renewcommand{\theequation}{\arabic{section}.\arabic{equation}}
\newcommand{\app}{\setcounter{section}{0}
\setcounter{equation}{0} \renewcommand{\thesection}{APPENDIX
\Alph{section}}\renewcommand{\theequation}{\Alph{section}.
\arabic{equation}}}
\usepackage{graphicx}
\usepackage{dcolumn}
\begin{document}
\title{Entropic Forms and Related Algebras}

\author{Antonio Maria Scarfone}
\email{antoniomaria.scarfone@cnr.it}

\affiliation{Istituto Sistemi Complessi (ISC--CNR) c/o Dipartimento di Scienza Applicata e Tecnologia-Politecnico di Torino, Corso Duca degli Abruzzi 24, Torino, I-10129, Italy}

\date{\today}
\begin{abstract}
Starting from a very general trace-form entropy, we introduce a pair of algebraic structures endowed by a generalized sum and a generalized product. These algebras form, respectively, two Abelian fields in the realm of the complex numbers isomorphic each other. We specify our results to several entropic forms related to distributions recurrently observed in social, economical, biological and physical systems including the stretched exponential, the power-law and the interpolating Bosons-Fermions distributions. Some potential applications in the study of complex systems are advanced.
\end{abstract}
 \maketitle

\sect{Introduction}

Complex systems have statistical properties that differ greatly from those of classical systems governed by the Boltzmann--Gibbs (BG) entropy \cite{CEJP,IJMP,MPLB,EPJST,CEJP1,AIP}. Often, the probability distribution observed in complex systems deviates from the Gibbs exponential one, showing asymptotic behavior characterized by stretched exponential \cite{Sornette}, power law \cite{Newman} or log-oscillating \cite{Zaslavsky} tails. To take into account these phenomena typically observed in social, economics, biological and physical systems \cite{Barabasi,Carbone,Carbone1,Sornette1,Blossey,Tsallis1}, several entropic forms \cite{Tsallis,Abe,Kaniadakis,Scarfone,Scarfone3,Anteneodo} have been introduced on the physical ground. They are derived by replacing the logarithmic function appearing in the BG entropy with its generalized version. In this way, the corresponding equilibrium distribution, obtained by maximizing the entropic form under appropriate constraints, assumes a generalized Gibbs expression with a deformed exponential instead of the \linebreak standard exponential.

In \cite{Kaniadakis,Scarfone1}, within the framework of the $\kappa$-deformed statistical mechanics, the idea of a generalized sum and product has been advanced with the purpose to extend some properties of standard logarithm and exponential to the corresponding $\kappa$-deformed functions. Shortly afterwards, generalized algebras and the related calculus have been proposed in the framework of the $q$-statistics based on Tsallis entropy \cite{Wang,Borges}. These algebras found several applications running from the number theory~\cite{Cardos,Lobao,Kaabouchi}, Laplace transformations \cite{Lenzi}, Fourier transformations \cite{Umarov,Jauregui1,Jauregui}, central limit theorem~\cite{Umarov1,Umarov2}, combinatorial analysis \cite{Niven2,Niven1,Suyari2,Suyari1,Oikonomou}, Gauss law of errors \cite{Scarfone2,Wada,Suyari}. On the physical ground, generalized algebras have been applied in statistical mechanics \cite{Lavagno} and statistical field theory \cite{Olemskoi}. The concept of generalized algebras has been employed constructively in \cite{Hanel,Hanel1} jointly to the introduction of the most general trace-form entropy compatible with the first three Khinchin axioms. In particular,~\cite{Hanel1} discussed the usefulness of a deformed product in studying the extensive property of generalized entropies, a question that will be further analyzed in the present work. Finally, in \cite{Tsallis2,Tsallis3}, generalized algebras has been used to account for statistical correlations originated from the fractal structures emerging in the phase space of interacting systems.

A question related to generalized algebras derived in the framework of statistical mechanics concerns the distributive propriety between deformed sum and deformed product. In the formalism of $\kappa$-statistics it has been shown in \cite{Scarfone1} that a distributive $\kappa$-sum and $\kappa$-product exist. These operations form an Abelian field in the domain of the real numbers. This means that $\kappa$-sum is associative, commutative, and admits the opposite and the neutral element in the field of the real numbers. In the same way, the $\kappa$-product is associative and commutative, and its inverse and identity exist and belong to the real~field.

Further, \cite{Kaniadakis} proposed a second Abelian algebra formed by a $\kappa$-sum and a $\kappa$-product different with the former. This algebra is not closed on the real field since the opposite element in the $\kappa$-sum is always a complex number. Thus, it forms an Abelian field only in the realm of the complex~numbers.

The situation appears more complicated in the framework of the $q$-statistics where a $q$-sum and a $q$-product have been introduced almost at the same time, but independently, in \cite{Wang,Borges}. These operations form separately two Abelian groups in $\mathbb{R}$ and in $\mathbb{R}^+$, respectively, although the $q$-product is not distributive with respect to the $q$-sum. Until today, the question related to the existence of a distributive algebra in the framework of $q$-statistics is still open as remarked in \cite{Tsallis4}. As will be shown, a possible solution to this problem is proposed in the following.

In this work we show that, starting from a well-defined complex function $g(z)$, a pair of Abelian fields defined in the complex plane and isomorphic to each other can be introduced. To make contact with statistical mechanics, we require that $g(z)$ is the analytical extension of a real and monotonic function $g(x)$, with $x\in U\subset(0,\,1)$, used in the definition of a generalized trace-form entropy. This is discussed next in Section 2, where we present the formalism and the notations used in this work. In Section~3, we introduce a systematic method to derive the algebraic structures related to a given generator $g(z)$. In Section 4, we detail our results to some relevant distributions emerging in the study of complex systems like the stretched exponential distribution, several power-law distributions and the interpolating Bose--Fermi quantum distribution. Potential applications concerning the Gauss law of errors and the study of the phase-space volume in an interacting system are investigated in Section 5. Final comments are reported in Section 6, and the proofs for the theorems stated in this paper are presented in the appendices.

\sect{Preliminary}

Let us introduce a real function $g(x):\,U\to V$, with $U\equiv\big(x_{\rm min},\,x_{\rm max}\big)\subset\mathbb{R}$ and   $V\equiv\big(y_{\rm min},\,y_{\rm max}\big)\subset\mathbb{R}$, where $x_{\rm min}\leq0$, $1\leq x_{\rm max}$, $y_{\rm min}\leq0\leq y_{\rm max}$. The function $g(x)$ is monotonic increasing in $U$
\begin{equation}
\frac{d\,g(x)}{d\,x}>0,\qquad\forall x\in U\label{con1}
\end{equation}
Without loss of generality, $g(x)$ can be normalized in $g(1)=0$. We require that any possible divergence of $g(x)$, as $x$ goes to zero, is mild enough to satisfy the condition
\begin{eqnarray}
\int\limits_0\limits^1g(x)\,dx>-\infty\label{e1}
\end{eqnarray}
Equation (\ref{con1}) assures the existence of a function $f(x):\,V\to U$, the inverse of $g(x)$, with
\begin{eqnarray}
f(g(x))=g(f(x))=x.
\end{eqnarray}
Clearly, Equation (\ref{con1}) induce the following conditions on $f(x)$:
\begin{eqnarray}
&&\frac{d\,f(x)}{d\,x}>0\label{con2a}
\end{eqnarray}
with $f(0)=1$ and
\begin{eqnarray}
\int\limits_{y_{\rm min}}\limits^0 f(x)\,dx<+\infty\label{con3}
\end{eqnarray}
From $g(x)$ we introduce a function $\Lambda(x)$ defined as
\begin{eqnarray}
\Lambda(x)={1\over x}\int\limits_0\limits^x g(y)\,dy-\Lambda_0,
\end{eqnarray}
with $\Lambda_0=\int_0^1g(x)\,dx$, which is strictly increasing, normalized in $\Lambda(1)=0$ and fulfills the relation
\begin{eqnarray}
{d\over d\,x}\left[x\,\Lambda(x)\right]=g(x)-\Lambda_0\label{deduced}
\end{eqnarray}
We define a generalized trace-form entropy according to
\begin{eqnarray}
S[p]=-\sum_{i=1}^Wp_i\,\Lambda(p_i)\label{i}
\end{eqnarray}
where $p=\{p_i\in [0,\,1]\}_{i=1,\,\ldots,\,W}$ is a set of normalized probability functions.

By integrating Equation (\ref{deduced}) on the interval $(0,\,p_i)$ and summing on the index $i$, we can rewrite the generalized entropy (\ref{i}) in the form
\begin{equation}
S[p]=-\sum_{i=1}^W\int\limits_0\limits^{p_i}g(x)\,dx+\Lambda_0\sum_{i=1}^Wp_i\label{ii}
\end{equation}
Accounting for the normalization of $p_i$, this expression coincides with the entropy-generating algorithm proposed in \cite{Abe1}.

Equation (\ref{i}) introduces a class of positive definite, symmetric, expandable, decisive, maximal, concave entropic forms. It includes the BG entropy as a particular case for $g(x)=\ln(x)$. Furthermore, as shown in \cite{Abe2}, Equation (\ref{i}) fulfills the Lesche stability condition, a fundamental property that should be satisfied by any well-established entropic function. Finally, the statistical mechanics theory based on these entropic forms satisfy the $H$-theorem when combined with the principle of   microscopic reversibility \cite{Abe1}.

Within the expression (\ref{i}), the search of the maximum of $S[p]$, constrained by $M+1$ linear relations
\begin{equation}
\sum_{i=1}^W\phi_j(x_i)\,p_i={\mathcal O}_j\label{const}
\end{equation}
with $j=0,\,1,\,\ldots M$, is realized throughout the following variational problem
\begin{equation}
{\delta\over\delta p_j}\left(S[p]-\sum_{j=0}^M\mu_j\sum_{i=1}^W\phi_j(x_i)\,p_i\right)=0\label{var}
\end{equation}
Quite often, the constraints $\phi_j(x)$ are given by means of momenta of a set of real numbers $x_i$ representing the possible outcomes of a certain physical observables $X$ occurring with probability $p_i$. For instance: ${\mathcal O}_0=1$, for $\phi_0(x_i)=1$, fixes the normalization; ${\mathcal O}_1\equiv\mu$, for $\phi_1(x_i)=x_i$, fixes the mean value of $X$; ${\mathcal O}_2\equiv\sigma^2$, for $\phi_2(x_i)=x_i^2$, fixes the variance of $X$
and so on.

Accounting for Equation (\ref{deduced}), the result of the problem (\ref{var}) reads
\begin{equation}
p_i\equiv p(x_i)=f(\tilde x_i)\label{fdi}
\end{equation}
with
\begin{equation}
\tilde x_i=\Lambda_0-\sum_{j=0}^M\mu_j\,\phi_j(x_i)
\end{equation}
Therefore, the distribution for the system described by the entropy (\ref{i}) is Gibbs-like, where the function $f(x)$ plays the role of a generalized exponential. The BG distribution is recovered for $f(x)=\exp(x)$.

It is worthy to note that generalized trace-form entropies and the related distributions by means of a variational principle {\em \`{a} la Jaynes} have been discussed in \cite{Hanel1,Hanel2,Thurner}. In particular, \cite{Hanel2,Thurner} have derived the most general trace-form entropy compatible with the first three Khinchin axioms that result to be a sub-family of the most general class (\ref{ii}).

Finally, we remark that Equations (\ref{con1})--(\ref{con3}) assure the existence of a well-defined entropic form and its distribution, obtained according to the maximal entropy principle. Nevertheless, these conditions are not strictly relevant to the definition of the algebraic structures discussed in the next section. Actually, some of these conditions can be relaxed if one is not interested in the entropic aspect of the problem but just interested in the link between the algebras and the pertinent distribution.

\sect{Generalized Algebras}\label{algebra}

In general, the function $g(x)$ and its inverse $f(x)$ can be extended analytically to the whole complex plane with the exception of a limited set of points $\{z_i\}$ corresponding to the branch of $g(z)$. In this way, let us consider the complex function $g(z)$, with $z\in\mathbb{C}_0\equiv\mathbb{C}-\{z_i\}$, and its inverse function $f(z)$, which is univocally defined in Riemann cut plane $\mathbb{C}_0$. They assume real values when $z\in U$ (resp. $z\in V$).

Starting from $g(z)$ and $f(z)$, we define a generalized sum and product forming a commutative and associative algebra in $\mathbb{C}_0$. They are given by
\begin{eqnarray}
&&z\oplus z'=f\big(\ln\big(\exp g(z)+\exp g(z')\big)\big)\label{s1}\\
&&z\otimes z'=f\big(g(z)+g(z')\big)\label{p1}
\end{eqnarray}
for any $z,\,z'\in \mathbb{C}_0$.\\

\noindent {\em Theorem 1.} The algebraic structure ${\mathcal A}\equiv\big(\mathbb{C}_0\times \mathbb{C}_0\to \mathbb{C}_0,\,\oplus,\,\otimes\big)$ form an Abelian field in $\mathbb{C}_0$.\\

This means that the generalized sum $\oplus$ fulfills:\\

\noindent (1). Commutativity: $z\oplus z'=z'\oplus z;$\\
(2). Associativity: $z\oplus(z'\oplus z'')=(z\oplus z')\oplus z'';$\\
(3). Neutral element: $z\oplus O=O\oplus z=z,$ where $O=\lim_{z\to-\infty}f(z);$\\
(4). Opposite: $z\oplus [-z]=[-z]\oplus z =O,$ where $[-z]=f(\ln(-\exp g(z)));$\\

\noindent the generalized product $\otimes$ fulfills:\\

\noindent (5). Commutativity: $z\otimes z'=z'\otimes z;$\\
(6). Associativity: $z\otimes(z'\otimes z'')=(z\otimes z')\otimes z'';$\\
(7). Identity element: $z\otimes 1=1\otimes z=z$;\\
(8). Inverse: $z\otimes [1/z]=[1/z]\otimes z =1,$ where $[1/z]=f(-g(z));$\\

\noindent and the product $\otimes$ is distributive w.r.t. the sum $\oplus$:\\

\noindent (9). Distributivity: $z\otimes(z'\oplus z'')=(z\otimes z')\oplus (z\otimes z'').$\\

\noindent (See Appendix A for a proof).\\

Remark that, in the domain of the real numbers, sum (\ref{s1}) is surely defined only if $V\equiv I\!\!R$, the opposite is always in $\mathbb{C}$ whilst the inverse is in $\mathbb{R}$ only if $y_{\rm min}=-y_{\rm max}$. Thus, limiting to the real field, the structure $(U\times U\to U,\,\oplus)$ form at least an Abelian semigroup.\\
The product $\otimes$ satisfies the relations
\begin{eqnarray}
&&f(z)\otimes f(z')=f(z+z')\label{f1}\\
&&g(z\otimes z')=g(z)+g(z')\label{g1}
\end{eqnarray}
which reflect the well-know properties of the standard exponential and logarithm: $\exp(z)\cdot\exp(z')=\exp(z+z')$ and $\ln(z\cdot z')=\ln(z)+\ln(z')$.\\ In the same way, posing $G(z)=\exp\big(g(z)\big)$ and $F(z)=f\big(\ln(z)\big)$, with $G\big(F(z)\big)=F\big(G(z)\big)=z$, the sum $\oplus$ fulfills the relations
\begin{eqnarray}
&&F(z)\oplus F(z')=F(z+z')\label{ff2}\\
&&G(z\oplus z')=G(z)+G(z')\label{gg2}
\end{eqnarray}
as dual of Equations (\ref{f1}) and (\ref{g1}), which now read
\begin{eqnarray}
&&F(z)\otimes F(z')=F(z\cdot z')\label{ff3}\\
&&G(z\otimes z')=G(z)\cdot G(z')\label{gg3}
\end{eqnarray}

A second pair of operations forming a commutative and associative algebra different from the previous one can be introduced by
\begin{eqnarray}
&&z\,\tilde\oplus\,z'=g\big(f(z)\cdot f(z')\big)\label{s2}\\
&&z\,\tilde\otimes\,z'=g\big(\exp\big(\ln f(z)\cdot\ln f(z')\big)\big)\label{p2}
\end{eqnarray}
for any $z,\,z'\in \mathbb{C}_0$.\\

\noindent {\em Theorem 2.} The algebraic structure $\widetilde{\mathcal A}\equiv\big(\mathbb{C}_0\times \mathbb{C}_0\to \mathbb{C}_0,\,\tilde\oplus,\,\tilde\otimes\big)$ forms an Abelian field in $\mathbb{C}_0$.\\

This means that the generalized sum $\tilde\oplus$ fulfills:\\

\noindent (1). Commutativity: $z\,\tilde\oplus\,z'=z'\,\tilde\oplus\,z;$\\
(2). Associativity: $z\,\tilde\oplus(z'\,\tilde\oplus\,z'')=(z\,\tilde\oplus\,z')\tilde\oplus\,z'';$\\
(3). Neutral element: $z\,\tilde\oplus\,0=0\,\tilde\oplus\,z=z;$\\
(4). Opposite: $z\,\tilde\oplus\,[-z]=[-z]\,\tilde\oplus\,z =0,$ where $[-z]=g(1/f(z));$\\

\noindent the generalized product $\tilde\otimes$
fulfills:\\

\noindent (5). Commutativity: $z\,\tilde\otimes\,z'=z'\,\tilde\otimes\,z;$\\
(6). Associativity: $z\,\tilde\otimes(z'\,\tilde\otimes\,z'')=(z\,\tilde\otimes\,z')\tilde\otimes\,z'';$\\
(7). Identity element: $z\,\tilde\otimes\,I=I\,\tilde\otimes\,z=z,$ where $I=g(e);$\\
(8). Inverse: $z\,\tilde\otimes\,[1/z]=[1/z]\,\tilde\otimes\,z=I,$ where $[1/z]=g\big(\exp(1/\ln f(z))\big);$\\

\noindent and the product $\tilde\otimes$ is distributive w.r.t. the sum $\tilde\oplus$:\\

\noindent (9). Distributivity: $z\,\tilde\otimes(z'\,\tilde\oplus\,z'')=(z\,\tilde\otimes\,z')\tilde\oplus\, (z\,\tilde\otimes\,z'').$\\

\noindent (See Appendix B for a proof).\\

This algebra is certain defined in $\mathbb{R}$ whenever $U\equiv\mathbb{R}^+$ and $V\equiv\mathbb{R}$.\\
The sum $\tilde\oplus$ satisfies the relations
\begin{eqnarray}
&&f(z\,\tilde\oplus\,z')=f(z)\cdot f(z')\label{f11}\\
&&g(z)\,\tilde\oplus\,g(z')=g(z\cdot z')\label{g11}
\end{eqnarray}
which, again, mimic the property of standard exponential and logarithm. Introducing the functions  $\tilde G(z)=g\big(\exp(z)\big)$ and $\tilde F(z)=\ln\big(f(z)\big)$, with $\tilde G\big(\tilde F(z)\big)=\tilde F\big(\tilde G(z)\big)=z$, the product $\tilde\otimes$ fulfills the~relations
\begin{eqnarray}
&&\tilde F(z\,\tilde\otimes\,z')=\tilde F(z)\cdot\tilde F(z')\label{f12}\\
&&\tilde G(z)\,\tilde\otimes\,\tilde G(z')=\tilde G(z\cdot z')\label{g12}
\end{eqnarray}
as dual of Equations (\ref{f11}) and (\ref{g11}), which now read as
\begin{eqnarray}
&&\tilde F(z\,\tilde\oplus\,z')=\tilde F(z)+\tilde F(z')\label{ff33}\\
&&\tilde G(z)\,\tilde\oplus\,\tilde G(z')=\tilde G(z+z')\label{gg33}
\end{eqnarray}

We observe that Equations (\ref{f11})--(\ref{g12}) follow from Equations (\ref{f1})--(\ref{gg2}), respectively, by posing
\begin{eqnarray}
&&\Big(\otimes\quad\rightarrow\quad\cdot\Big),\qquad\Big(+\quad\rightarrow\quad\tilde\oplus\Big)\\
&&\Big(\oplus\quad\rightarrow\quad\cdot\Big),\qquad\Big(+\quad\rightarrow\quad\tilde\otimes\Big)
\end{eqnarray}
and vice versa.\\
This correspondence can be summarized in the following\\

\noindent {\em Theorem 3.} The algebraic structures ${\mathcal A}$ and $\widetilde{\mathcal A}$ are homomorphic.\\

The homomorphism is established by the relations:
\begin{eqnarray}
&&(z\oplus z')_g=\big(w\,\tilde\oplus\,w'\big)^f\\
&&(z\otimes z')_g=\big(w\,\tilde\otimes\,w'\big)^f
\end{eqnarray}
with $z_g=\exp\big(g(z)\big)$ and $z^f=\ln\big(f(z)\big)$, where $w$ and $w'$ are solutions of equation $w^f=z_g$. (See Appendix C for a proof).

\sect{Examples}

\subsect{Stretched-Exponential Distribution}

As a first example let us consider the following entropic form \cite{Anteneodo}
\begin{equation}
S_\gamma[p]=\sum_{i=1}^W\Gamma\left({1\over\gamma}+1,\,-\ln p_i\right)-\Gamma\left({1\over\gamma}+1\right)\sum_{i=1}^Wp_i\label{strech}
\end{equation}
where $\Gamma(u,\,x)=\int_x^\infty t^{u-1}\,\exp(-t)\,dt$ is the incomplete gamma function of the second kind and   $\Gamma(u)\equiv\Gamma(u,\,0)$ is the gamma function \cite{Gradshteyn}. Equation (\ref{strech}) follows from Equation (\ref{ii}) with
\begin{equation}
g(x)=-\Big(\ln {1\over x}\Big)^{1/\gamma}
\end{equation}
with $\gamma>0$. This function assumes real values for $x\subset U\in\big(0,\,1\big)$ with $y\subset V\in\big(-\infty,\,0\big)$ and the distribution takes the form of a stretched exponential
\begin{equation}
p_i=\exp\Big(-\tilde x_i^{\,\,\gamma}\Big)\label{strech1}
\end{equation}
The first algebraic structure related to entropy (\ref{strech}) is generated by the operations
\begin{eqnarray}
&&z\oplus z'=\left[\exp\left(\ln{E_\gamma(z)\,E_\gamma(z')\over E_\gamma(z)+E_\gamma(z')}\right)^\gamma\right]^{-1}\\
&&z\otimes z'=\left[\exp\big(\ln(E_\gamma(z)\,E_\gamma(z'))\big)^\gamma\right]^{-1}
\end{eqnarray}
where $E_\gamma(z)=\exp(-\ln z)^{1/\gamma}$, the opposite $[-z]=1/E_{1/\gamma}(-E_\gamma(z))\in\mathbb{C}$ and the inverse $[1/z]=z^I$, with $I=(-1)^\gamma$, in general, belonging to $\mathbb{C}$.\\ The second algebraic structure $\widetilde{\cal A}$ is given by
\begin{eqnarray}
&&z\,\tilde\oplus\,z'=\left(z^\gamma+z'^\gamma\right)^{1/\gamma}\\
&&z\,\tilde\otimes\,z'=I\,z\,z'
\end{eqnarray}
where $I=(-1)^{1+1/\gamma}$ and $[-z]=-I\,z$ are, in general, in $\mathbb{C}$.

\subsect{Power-Law Like Distributions}

A next example is given by the Sharma--Taneja--Mittal (STM) entropy \cite{Taneja1,Taneja2,Mittal}
\begin{equation}
S_{\kappa,\,r}[p]=-\sum_{i=1}^W{p_i^{1+r+\kappa}-p_i^{1+r-\kappa}\over2\,\kappa}\label{stm}
\end{equation}
with $-|\kappa|\leq r\leq|\kappa|$ if $0\leq|\kappa|<1/2$ and $|\kappa|-1\leq r\leq1-|\kappa|$ if $1/2\leq|\kappa|<1$.

This entropic form, reconsidered on the physical ground in \cite{Scarfone3,Scarfone1}, can be obtained from   Equation (\ref{ii}) by posing
\begin{equation}
g(x)=\lambda\,\ln_{\{\kappa,\,r\}}\left({x\over\alpha}\right)-1\label{gkr}
\end{equation}
where
\begin{equation}
\ln_{\{\kappa,\,r\}}(x)={x^{r+\kappa}-x^{r-\kappa}\over2\,\kappa}\label{krl}
\end{equation}
is the deformed logarithm. The existence of $\exp_{\{\kappa,\,r\}}(x)$, the inverse function of $\ln_{\{\kappa,\,r\}}(x)$, follows from its monotonicity in $\mathbb{R}$ although an explicit expression, in general, cannot be given. The two constants $\alpha$ and $\lambda$ are given by
\begin{eqnarray}
\alpha=\left(\frac{1+r-\kappa}{1+r+\kappa}\right)^{1/2\,\kappa},\qquad\lambda=\frac{\left(1+r-\kappa\right)}{\mbox{\raisebox{-1mm}
{$\left(1+r+\kappa\right)$}}}
^{\scriptscriptstyle{(r+\kappa)/2\,\kappa}}_ {\mbox{\raisebox{3.5mm}{$\scriptscriptstyle{(r-\kappa)/2\,\kappa}$}}}\label{l}
\end{eqnarray}
and are related by $(1+r \pm \kappa)\,\alpha^{r \pm
\kappa}=\lambda$ and $1/\lambda=\ln_{\{\kappa,\,r\}}\left(1/\alpha\right)$. Therefore, function (\ref{gkr}) is correctly normalized. The equilibrium distribution related to entropy (\ref{stm}) is
\begin{eqnarray}
p(\tilde x_i)=\alpha\,\exp_{\{\kappa,r\}}\left(-{\tilde x_i\over\lambda}\right)\label{stmd}
\end{eqnarray}
and the algebras can be obtained from Equations (\ref{s1})--(\ref{p1}) and Equations (\ref{s2})--(\ref{p2}).\\
Hereinafter, we consider some particular cases:\\

\noindent \emph{$q$-Distribution}\\

We look at the following entropic form
\begin{equation}
S_{2-q}[p]=\sum_{i=1}^W{p_i^{2-q}-p_i\over q-1}\label{ts}
\end{equation}
with $0<q<2$, known as Tsallis entropy \cite{Tsallis} in the $(2-q)$-formalism \cite{Wada1}. It belongs to the   STM-family with $r=|\kappa|$ and $q=1-2\,|\kappa|$ and its equilibrium distribution reads
\begin{eqnarray}
p_i=\alpha_q\,\left[1-(1-q)\,\tilde x_i\right]^{1\over 1-q}\label{tt}
\end{eqnarray}
with $\alpha_q=(2-q)^{1/(q-1)}$ and $\lambda_q=1$.\\
Entropy (\ref{ts}) follows from Equation (\ref{ii}) by posing
\begin{equation}
g(x)={x^{1-q}-1\over a_q}\label{gq}
\end{equation}
with $a_q=(1-q)/(2-q)$. This function takes real values for $U\equiv\mathbb{R}^+$ with $V\equiv\big(-1/a_q,\,+\infty\big)$ for $0<q<1$ and $V\equiv\big(-\infty,\,-1/a_q\big)$ for $1<q<2$.\\
The first algebraic structure related to the $q$-distribution is generated by the operations
\begin{eqnarray}
&&z\oplus z'=\left[{z^{1-q}+z'^{1-q}\over2}+a_q\,\ln\Big(2\,\cosh\Big( {z^{1-q}-z'^{1-q}\over 2\,a_q}\Big)\Big)\right]^{1\over1-q}\label{qsum}\\
&&z\otimes z'=\big(z^{1-q}+z'^{1-q}-1\big)^{1\over1-q}\label{qprod}
\end{eqnarray}
where $O\in\mathbb{C}$ for $0<q<1$, $O\equiv 0$ for $1<q<2$ and $[1/z]=(2-z^{1-q})^{1/(1-q)}$.\\
The second algebraic structure follows from
\begin{eqnarray}
&&z\,\tilde\oplus\,z'=z+z'+a_q\,z\,z'\label{qsum1}\\
&&z\,\tilde\otimes\,z'={1\over a_q}\left[\exp\left({1\over1-q}\,\ln(1+a_q\,z)\,\ln(1+a_q\,z')\right)-1\right]\label{qprod1}
\end{eqnarray}
with $[-z]=-z/(1+a_q\,z)$, $I=(e^{1-q}-1)/a_q$ and $[1/ z]=(e^{(1-q)^2/\ln(1+a_q\,z)}-1)/a_q$.\\

As already stated, $q$-product (\ref{qprod}) and $q$-sum (\ref{qsum1}) have been introduced in literature in \cite{Borges,Wang} and recently they were widely employed in the framework of $q$-statistics \cite{Lenzi,Umarov,Umarov1,Umarov2}. It is now clear why operations (\ref{qprod}) and (\ref{qsum1}) do not fulfil the distributive properties $z\,\tilde\oplus\,(z'\otimes z'')\not=z\,\tilde\oplus\,z'\otimes z\,\tilde\oplus\,z''$ since they belong to two different algebraic structures. \\
Often, $q$-algebra is applied directly to the functions
\begin{eqnarray}
\ln_q(z)={z^{1-q}-1\over1-q}
\end{eqnarray}
and
\begin{eqnarray}
\exp_q(z)=[1+(1-q)z]^{1/(1-q)}
\end{eqnarray}
They fulfill the relations
\begin{eqnarray}
&&\ln_q(z\otimes_q z')=\ln_q(z)+\ln_q(z')\\
&&\exp_q(z+z')=\exp_q(z)\otimes_q\exp_q(z')
\end{eqnarray}
and
\begin{eqnarray}
&&\ln_q(z\cdot z')=\ln_q(z)\,\tilde\oplus_q\,\ln_q(z')\\
&&\exp_q(z\,\tilde\oplus_q\, z')=\exp_q(z)\cdot\exp_q(z')
\end{eqnarray}
Then, it could be useful to rewrite Equations (\ref{qsum})--(\ref{qprod1}) with respect to $\ln_q(z)$ and $\exp_q(z)$ to make contact with the existent literature. They are
\begin{eqnarray}
&&z\oplus_q z'=\exp_q\left(\ln\left(\exp\ln_q(z)+\exp\ln_q(z')\right)\right)\\
&&z\otimes_q z'=\big(z^{1-q}+z'^{1-q}-1\big)^{1\over1-q}
\end{eqnarray}
and
\begin{eqnarray}
&&z\,\tilde\oplus_q\,z'=z+z'+(1-q)\,z\,z'\\
&&z\,\tilde\otimes_q\,z'=\ln_q\left(\exp\left(\ln\exp_q(z)\cdot\ln\exp_q(z')\right)\right)
\end{eqnarray}

\noindent \emph{Student's $t$-Distribution}\\

In statistics the $q$-distribution recovers, for special values of $q$, the Student' $t$-distribution. In fact, passing to the continuum, entropy (\ref{ts}) can be written in
\begin{equation}
S_\nu[p]={1\over\nu}\int\limits_{-\infty}\limits^\infty\left(p(x)^{1-\nu}-p(x)\right)\,dx\label{student}
\end{equation}
with $\nu=q-1$.

Maximization of Equation (\ref{student}), under the constraints of normalization and variance   $\sigma^2=(2-\nu)/(2-3\,\nu)$, with $0<\nu<2/3$, gives the Student's $t$-distribution
\begin{eqnarray}
p(x)=\sqrt{\nu\over\pi\,(2-\nu)}\,{\Gamma\left({1\over\nu}\right)\over\Gamma\left({1\over\nu}-{1\over2}\right)}\,
\left(1+{\nu\over2-\nu}\,x^2\right)^{-1/\nu}\label{st}
\end{eqnarray}
which can be rewritten in the form
\begin{eqnarray}
p(\tilde x)=\left(1+\nu\,\tilde x\right)^{-1/\nu}
\end{eqnarray}
with $\tilde x=(1+\mu_0+\mu_2\,x^2)/(1-\nu)$. In this case, the related algebras follow readily from Equations (\ref{qsum})--(\ref{qprod1}).\\

\noindent \emph{Cauchy--Lorentz distribution}\\

In the $\nu\to1$ limit, Equation (\ref{st}) reduces to the Cauchy--Lorentz distribution
\begin{eqnarray}
p(x)={1\over\pi}\,{1\over 1+x^2}\label{lorentz}
\end{eqnarray}
here rewritten in
\begin{eqnarray}
p(\tilde x)={1\over 1-\tilde x}
\end{eqnarray}
with $\tilde x=1-\pi\,(1+x^2)<0$. Although Equation (\ref{lorentz}) cannot be obtained from the variational   problem (\ref{var}) since, as known, the second moment of Cauchy distribution is not defined, we can still derive the corresponding algebra starting from the distribution, by posing
\begin{equation}
g(z)=1-{1\over z}\label{glorentz}
\end{equation}
in the spirit of Equation (\ref{fdi}).\\
We obtain
\begin{eqnarray}
&&z\oplus z'=\left[{z+z'\over 2\,z\,z'}-\ln\left(2\,\cosh\left({z-z'\over 2\,z\,z'}\right)\right)\right]^{-1}\\
&&z\otimes z'={z\,z'\over z+z'-z\,z'}
\end{eqnarray}
and
\begin{eqnarray}
&&z\,\tilde\oplus\,z'=z+z'-z\,z'\\
&&z\,\tilde\otimes\,z'=1-\exp\big(-\ln(1-z)\,\ln(1-z')\big)
\end{eqnarray}
We observe that Equation (\ref{glorentz}) does not fulfill condition (\ref{e1}). However, in this case, since an entropic form related to the Cauchy distribution is not known, at least in the sense of a variational principle, the generator of the algebras is related just to the distribution.\\

\noindent \emph{Zipf--Pareto Distribution}\\

Starting from Equation (\ref{krl}) we can derive the asymptotic behavior of the distribution functions, since
\begin{eqnarray}
\exp_{\{\kappa,r\}}(-x)=1/\exp_{\{\kappa,-r\}}(x)\sim x^{-1/(k-r)}
\end{eqnarray}
which holds for $x\gg1$. Therefore, Equation (\ref{stmd}) asymptotically behaves like the Zipf--Pareto distribution
\begin{eqnarray}
p(x)\sim x^s\label{pareto}
\end{eqnarray}
with $s=1/(r-k)<0$.

Formally, we can obtain the corresponding algebras, in the limit of large values of $z$ and $z'$, starting from the algebras related to the STM entropy. The corresponding operations are:
\begin{eqnarray}
&&z\oplus z'\sim\left[\ln\left(\exp z^{1/s}+\exp z'^{1/s}\right)\right]^s\label{zp1}\\
&&z\otimes z'\sim(z^{1/s}+z'^{1/s})^s
\end{eqnarray}
and
\begin{eqnarray}
&&z\,\tilde\oplus\,z'\sim z\,z'\\
&&z\,\tilde\otimes\,z'\sim\exp\left(s\,\ln z\,\ln z'\right)\label{zp4}
\end{eqnarray}
respectively.

Remark that these operations are obtained in the limit of large values starting from the operations obtained in the STM formalism. Consequently, Equations (\ref{zp1})--(\ref{zp4}) cannot be considered as a continuous deformation of the standard operations of sum and product, although the algebras formed by these operations are still commutative, associative and distributed. (Standard algebra are recovered in the $\gamma\to1$ limit for the stretched exponential and in the $(\kappa,\,r)\to(0,\,0)$ limit for STM family.)

\subsection{Interpolating Fermi--Dirac and Bose--Einstein Distribution}

As a final example, let us consider the interpolating Bosons-Fermions distribution. Following \cite{Quarati}, we start from the equilibrium distribution
\begin{equation}
p(\tilde x_i)={1\over\exp(\tilde x_i)-\kappa}\label{bfd}
\end{equation}
where $\kappa\in(-1,\,1)$ is the interpolating parameter. In the $\kappa\to\pm1$ limit we obtain the well-known Bose--Einstein and Fermi--Dirac statistics. The Boltzmann--Gibbs distribution is also included for $\kappa=0$.

Equation (\ref{bfd}) suggests considering the following generating function
\begin{eqnarray}
g(x)=\ln{1+\kappa\,x\over x}
\end{eqnarray}
Unfortunately, this function does not meet the correct normalization $f(0)=1$. This happens because in quantum many-body statistics, the normalization factor, the fugacity, is accounted inside to the exponential term and not as an overall factor like in the formalism presented in these pages.

To obtain the right normalization we scale opportunely the axes origin. In this way, the generator~becomes
\begin{eqnarray}
&&g(x)=\ln{(1+\kappa)\,x\over 1+\kappa\,x}\label{bfg}
\end{eqnarray}
and the corresponding entropic form reads
\begin{equation}
S_\kappa[p]=-\sum_{i=1}^W\left[p_i\ln p_i-{1+\kappa\,p_i\over\kappa}\,\ln(1+\kappa\,p_i)\right]-{1+\kappa\over\kappa}\,\ln(1+\kappa)\label{bfe}
\end{equation}
which coincides with the one proposed in \cite{Quarati}, modulo the additive constant factor.

Function (\ref{bfg}) assumes real values for $U\in\big(-\infty,\,-1/\kappa\big)\cup\big(0,\,+\infty\big)$ when $\kappa>0$ and $U\in\big(0,\,-1/\kappa\big)$ when $\kappa<0$, with $V\in\big(-\infty,\,+\infty\big)$.

The first set of operations are given by
\begin{eqnarray}
&&z\oplus z'={z+z'+2\,\kappa\,z\,z'\over1+\kappa^2\,z\,z'}\\
&&z\otimes z'={(1+\kappa)\,z\,z'\over1+\kappa\,(z+z'-z\,z')} \label{bfp1}
\end{eqnarray}
with $[-z]=-z/(1+2\,\kappa\,z)$, $[1/z]=(1+\kappa\,z)/[(1+2\,\kappa)\,z-\kappa]$, while the latter are
\begin{eqnarray}
&&z\,\tilde\oplus\,z'=-\ln\big(\kappa-\kappa\,\exp(-z)-\kappa\,\exp(-z'))
+(1+\kappa)\exp(-z-z')\big) \ \\
&&z\,\tilde\otimes\,z'=\ln(1+\kappa)-\ln\big(\kappa+\exp(-h(z)\,h(z'))\big) \ \label{bfp2}
\end{eqnarray}
where $h(z)=\ln\big(1+\kappa-\kappa\,\exp(z)\big)-z$, $[-z]=-z+\ln(1+\kappa-\kappa\,\exp(z))-\ln(1-\kappa+\kappa\,\exp(-z))$, $I=\ln\big((1+\kappa)\,e/(1+\kappa\,e)\big)$ and $[1/z]=\ln(1+\kappa)-\ln(\kappa+\exp(1/h(z)))$.

\sect{Applications}

\subsection{Gauss Law of Error}

One of the most known and ubiquitous distributions in nature is surely the Gauss law of error, also known as normal distribution. Its universality is well explained by the central limit theorem, which establishes its role of ``attractor'' in the space of distribution in the limit of a large number of statistically independent and identically distributed events.

Recently \cite{Umarov,Umarov1}, a possible generalization of the central limit theorem has been proposed in order to justify the recurrence of non-Gaussian distributions \cite{Scarfone2,Wada,Suyari} in the limit of a large number of statistically dependent events.

In the following, we reconsider the derivation of Gauss law for generalized correlated measurements, where correlations are accounted for by means of a deformed product.

Let us consider $n$ replicate measurements of a given observable. Let $x_i\in \Re$, with $i=1,\ldots,\,n$, the experimental values obtained. These are distributed around the ``exact'' value $x_{\rm e}$, although unknown, so that $\epsilon_i=|x_i-x_{\rm e}|$ represents the ``error'' or uncertainty in each measurement. The quantities $\epsilon_i$ can be treated as the outcomes of a random variable with a discrete probability distribution function $p_i=p(\epsilon_i)$.

From the experimental measurement, one obtains the mean value
\begin{equation}
x^\ast={1\over n}\,(x_1+x_2+\ldots+x_n)\label{valmed}
\end{equation}
and, according to the law of large numbers \cite{Feller}
\begin{equation}
x_{\rm e}=\lim_{n\to\infty}x^\ast
\end{equation}
We suppose that in a series of $n$ subsequent measurements, each event is not necessarily statistically independent and correlations among events can be taken into account by means of a generalized product.

Following the maximal likelihood principle, we introduce the quantity
\begin{eqnarray}
{\cal L}(x_{\rm e})=p(\epsilon_1)\otimes p(\epsilon_2)\otimes\ldots\otimes p(\epsilon_n)\label{like}
\end{eqnarray}
as a function of $x_{\rm e}$, since
\begin{eqnarray}
\epsilon_i=|x_i-x_{\rm e}|
\end{eqnarray}
and $p(\epsilon_i)$ can be obtained by maximizing Equation (\ref{like}) under the condition $x^\ast=x_{\rm e}$.

However, it is more convenient to evaluate the extremum of $g({\cal L}(x))$; being $g(x)$ a monotonic increasing function, its maximum coincides with that of ${\cal L}(x)$. Thus, from Equation (\ref{g1}), we have
\begin{equation}
g\Big({\cal L}(x_{\rm e})\Big)=\sum_{i=1}^ng
\Big(p(\epsilon_i)\Big)\label{ll}
\end{equation}
and extremal points are obtained from equation
\begin{equation}
\sum_{i=1}^n{d\over d\,y_i}\,g(p(y_i))\Big|_{y_i=x_i-x^\ast}=0\label{eq1}
\end{equation}
On the other hand, the algebraic equation $\sum_i\phi(y_i)=0$, constrained by the condition $\sum_iy_i=0$, admits the solution $\phi(y)=a\,y$ for a given real constant $a$ \cite{Suyari} so that, accounting for
\begin{equation}
\sum_{i=1}^ny_i\equiv\sum_{i=1}^n(x_i-x^\ast)=0
\end{equation}
we get
\begin{equation}
{d\over d\,y}\,g(p(y))=-2\,\beta\,y
\end{equation}
with $a\equiv-2\,\beta$, a constant.

After integration, the distribution assumes the final form
\begin{equation}
p(x_i)=f\Big(-\mu-\beta\,(x_i-x)^2\Big)\label{sol}
\end{equation}
where the integration constant $\mu$ is fixed by the normalization.

This function is a generalized Gaussian distribution provided that $\beta$ is positive definite.

Furthermore, the positivity of $\beta$ is required to guarantee the maximum of ${\cal L}(x)$ (and not a minimum) at $x\equiv x_{\rm e}$. In fact, using Equations (\ref{ll}) and (\ref{sol}), we obtain
\begin{equation}
{\cal L}(x_{\rm e})=f\Bigg(-n\,\mu
-\beta\sum_{i=1}^n(x_i-x_{\rm e})^2\Bigg)
\end{equation}
and maximum at $x\equiv x_{\rm e}$ requires
\begin{equation}
{d^2\,{\cal L}(x)\over d\,x^2}\Bigg|_{x=x^\ast}=\left[{d^2\,f(y)\over d\,y^2}
\,\left({d\,y\over d\,x}\right)^2+{d\,f(y)\over d\,y}\,{d^2\,y\over d\,x^2}\right]_{x=x^\ast}<0\label{ine}
\end{equation}
with $y=-n\,\mu-\beta\sum_i(x_i-x)^2$.

Positivity of $\beta$ follows from $d\,f(x)/d\,x>0$ and observing that
\begin{equation}
{d\,y\over d\,x}\Big|_{x=x^\ast}=0
\end{equation}
and
\begin{equation}
{d^2\,y\over d\,x^2}\Big|_{x=x^\ast}=-2\,\beta
\end{equation}

\subsection{Extensivity of Microcanonical Entropy}

In the microcanonical picture, the macrostate ``visited'' by a single particle is formed by a collection of $W$ microstates all having the probability $p=1/W$ to be inspected. In this simple case, $\Gamma$-space and $\mu$-space coincide and BG entropy reduces to the Boltzmann--Planck entropy
\begin{eqnarray}
S_{\rm B}=\ln W\label{BP}
\end{eqnarray}
for a set of $W$ identically distributed elements.

As known, Equation (\ref{BP}) becomes an extensive quantity when the system is composed by $N$ identically distributed and uncorrelated particles, such that
\begin{eqnarray}
p_N={1\over W^N}\equiv\left({1\over W}\right)^N=(p)^N \label{fac}
\end{eqnarray}
and the probability in the $\Gamma$-space factorizes into the product of the respective marginal probabilities of the corresponding $\mu$-space. In this situation entropy (\ref{BP}) scales according to $S_{\rm B}(N)=N\,S_{\rm B}(1)$.

Clearly, the validity of Equation (\ref{fac}) is related to the assumption of statistical independence between the particles of the system. Otherwise, if the system is globally correlated, this property could be no longer valid and other entropic forms may become extensive depending on the nature of the correlations among the particles. In \cite{Tsallis2,Tsallis3}, it has been suggested that, in order to restore the extensive property of entropy, the $q$-product introduced in \cite{Borges,Wang} can be used to describe the possible correlations in a system governed by the $q$-entropy.

Actually, this is a general statement in the microcanonical picture. In fact, let us consider a collection of $W$ identically distributed events with equal probability $p_i=1/W$, $(i=1,\ldots,\,W)$.   Equation (\ref{i}) becomes
\begin{eqnarray}
S[W]=-\sum_{i=1}^W{1\over W}\,\Lambda\left({1\over W}\right)=-\Lambda\left({1\over W}\right)\label{gBP}
\end{eqnarray}
for a given monotonic increasing function $\Lambda(x)$, which is a generalized version of Boltzmann--Planck entropy.

By identifying the functions $\Lambda(x)$ and its inverse ${\cal E}(x)$ with the algebraic generators $f(x)$ and $g(x)$, respectively, we can introduce the product $\otimes$ according to
\begin{eqnarray}
\Lambda(x\,\otimes\,y)=\Lambda(x)+\Lambda(y)
\end{eqnarray}
We postulate the following composition rule in the $\Gamma$-space
\begin{eqnarray}
W^{\rm AB}=\left({1\over W^{\rm A}}\otimes{1\over W^{\rm B}}\right)^{-1}\label{composition}
\end{eqnarray}
Thus, entropy (\ref{gBP}) becomes additive
\begin{eqnarray}
\nonumber S[W^{\rm AB}]&=&-\Lambda\left({1\over W^{\rm AB}}\right)\\
\nonumber
&=&-\Lambda\left({1\over W^{\rm A}}\otimes{1\over W^{\rm B}}\right)\\
\nonumber
&=&-\Lambda\left({1\over W^{\rm A}}\right)-\Lambda\left({1\over W^{\rm B}}\right)\\
&=&S[W^{\rm A}]+S[W^{\rm B}]
\end{eqnarray}
and, more in general, extensive
\begin{eqnarray}
\nonumber S[W^N]&=&-\Lambda\left({1\over W^N}\right)\\
\nonumber
&=&-\Lambda\left(\left({1\over W}\right)^{\otimes\,N}\right)\\
\nonumber
&=&-N\,\Lambda\left({1\over W}\right)\\
&=&N\,S[W]
\end{eqnarray}
where $x^{\otimes\,N}=x\,\otimes\,x\otimes\,x\,\otimes\ldots\otimes\,x$, $N$ times.\\
As an explicit example, let us consider the microcanonical version of entropy (\ref{ts}) that is
\begin{eqnarray}
&&S_{2-q}[W]=\ln_q(W)\label{bq}
\end{eqnarray}
From the corresponding algebra we obtain
\begin{eqnarray}
W^{\rm AB}=\left[\left(W^{\rm A}\right)^{q-1}+\left(W^{\rm B}\right)^{q-1}-1\right]^{1/(q-1)}
\end{eqnarray}
and
\begin{eqnarray}
W^N=\left(N\,W^{q-1}-N+1\right)^{1/(q-1)}
\end{eqnarray}
making entropy (\ref{bq}) additive and extensive. Analogue relations have been already derived in \cite{Tsallis3}.

Another handling example is given by $\kappa$-entropy \cite{Kaniadakis}
\begin{equation}
S_\kappa[p]=-\sum_{i=1}^W{p_i^{1+\kappa}-p_i^{1-\kappa}\over 2\,\kappa}\label{kappa}
\end{equation}
with $|\kappa|<1$, which is a one parameter subfamily of the STM entropy (\ref{stm}), with $r=0$.

In the microcanonical picture entropy (\ref{kappa}) becomes \cite{Scarfone3}
\begin{eqnarray}
S[W]=\ln_{\{\kappa\}}(W)
\end{eqnarray}
where
\begin{eqnarray}
\ln_{\{\kappa\}}(x)={x^\kappa-x^{-\kappa}\over2\,\kappa}
\end{eqnarray}
is the $\kappa$-logarithm \cite{Kaniadakis}.

The algebraic operations obtained from the function $\ln_\kappa(z)$ are, respectively
\begin{eqnarray}
&&z\oplus z'=\exp\left({1\over\kappa}\,{\rm arcsinh}\left(h_+(z,\, z')+\kappa\,\ln\left(2\,\cosh\left({1\over\kappa}\,h_-(z,\,z')\right)\right)\right)\right)\label{k1} \\
&&z\otimes z'=\exp\left({1\over\kappa}\,{\rm arcsinh}\left(2\,h_+(z,\,z')\right)\right)
\end{eqnarray}
with
\begin{eqnarray}
h_\pm(z,\,z')={1\over4}\,\left[z^{\kappa}\pm z'^{\kappa}-(z^{-\kappa}\pm z'^{-\kappa})\right]
\end{eqnarray}
and
\begin{eqnarray}
&&z\,\tilde\oplus\,z'=z\,\sqrt{1+\kappa^2\,z'^2}+z'\,\sqrt{1+\kappa^2\,z^2},\\
&&z\,\tilde\otimes\,z'={1\over\kappa}\,\sinh\left({1\over\kappa}\,{\rm arcsinh}(\kappa\,z)\,{\rm arcsinh}(\kappa\,z')\right)\label{k2}
\end{eqnarray}
Remark that algebra $(\mathbb{R}^+\times\mathbb{R}^+\to\mathbb{R}^+,\,\oplus,\,\otimes)$ is a unitary semi-ring. Differently, algebra $(\mathbb{R}\times\mathbb{R}\to\mathbb{R},\,\tilde\oplus,\,\tilde\otimes)$
forms an Abelian field.

Using Equation (\ref{k1}) we obtain
\begin{eqnarray}
W^{\rm AB}=\Big(2\,h_+(W^{\rm A},\,W^{\rm B})+\sqrt{1+4\,h_+(W^{\rm A},\,W^{\rm B})^2}\Big)^{1/\kappa}
\end{eqnarray}
and
\begin{eqnarray}
W^N=\Big[N\,\kappa\,\Lambda_\kappa(W)+\sqrt{1+(N\,\kappa\,\Lambda_\kappa(W))^2}\Big]^{1/\kappa}
\end{eqnarray}
In the limit of $W\to\infty$, we get
\begin{eqnarray}
W^N\quad\sim_{_{_{\hspace{-4mm}W\to\infty}}} N^{1/\kappa}\,W\label{scale}
\end{eqnarray}
Thus, $\kappa$-entropy is asymptotically extensive whenever the volume of the phase space scales   according to $N^{1/\kappa}$.

We remark that operations (\ref{k1})--(\ref{k2}) coincide with the ones introduced in \cite{Kaniadakis}. There, author introduces $\kappa$-algebras in the real field starting from the relations
\begin{eqnarray}
&&(x\oplus y)_{\{\kappa\}}=x_{\{\kappa\}}+y_{\{\kappa\}}\label{ka1}\\
&&(x\otimes y)_{\{\kappa\}}=x_{\{\kappa\}}\cdot y_{\{\kappa\}}
\end{eqnarray}
and
\begin{eqnarray}
&&(x\oplus y)^{\{\kappa\}}=x^{\{\kappa\}}+y^{
\{\kappa\}}\\
&&(x\otimes y)^{\{\kappa\}}=x^{\{\kappa\}}\cdot y^{\{\kappa\}}\label{ka2}
\end{eqnarray}
where the $\kappa$-numbers $x_{\{\kappa\}}$ and $x^{\{\kappa\}}$ are defined in
\begin{eqnarray}
&&x_{\{\kappa\}}={1\over\kappa}\,{\rm arcsinh}\,h(\kappa\,x)\\
&&x^{\{\kappa\}}={1\over\kappa}\,h^{-1}\big(\sinh(\kappa\,x)\big)
\end{eqnarray}
for a given generator $h(x)$. It is easy to verify as Equations (\ref{ka1})--(\ref{ka2}) are substantially equivalent to Equations (\ref{f12}), (\ref{ff33}) and (\ref{ff2}), (\ref{ff3}), respectively, where the correspondence arises from
\begin{eqnarray}
&&g(z)=\ln z^{\{\kappa\}}\\
&&f(z)=\exp z_{\{\kappa\}}
\end{eqnarray}
Finally, it is worthy to note that, a scaling relation similar to Equation (\ref{scale}) has been considered in \cite{Kaniadakis3}, where it has been  discussed a possible application of $\kappa$-statistics to the Bekenstein--Hawking are law in the framework of black holes thermostatistics.
\sect{Conclusions}

In the picture of a statistical mechanics based on a generalized trace-form entropy, we derived two algebraic structures whose operations of sum and product generalize the well-known proprieties of sum and product of ordinary logarithm and exponential. These deformed functions are often used to study statistical mechanics properties of systems characterized by non-Gibbsian distributions. We stated a link between the formalism of generalized algebras and that used in the formulation of generalized statistical mechanics since, as shown, the former are derived starting from a very general expression of a trace-form entropy. In this way, one can introduce a pair of algebraic structures useful for the mathematical manipulation of the emerging statistical-mechanics theory.

We presented several examples to illustrate how the method works. For each of them, we have listed the main quantities like the neutral element for the sum and the unity for the product. Often, these quantities take complex values, \emph{i.e}., algebras form Abelian field in $\mathbb{C}$, since, in general, they are not closed in $\mathbb{R}$. However, complex algebra turns out to be relevant in several aspects, such as in the definition of generalized a $q$-Fourier transformation \cite{Umarov,Jauregui1}, given by
\begin{eqnarray}
F_q[f](\xi)=\int_{{\rm supp }f}\exp_q(i\,\chi\,\xi)\otimes_q f(\chi)\,d\chi
\end{eqnarray}
and other \cite{Jauregui,Umarov1,Umarov2}. Furthermore, applications of statistical mechanics to the study of non-equilibrium phenomena, like damping harmonic oscillator or nuclear decay, the frequencies of oscillation or the nuclear energy level are assumed to be complex quantities with the imaginary part related to the damping effect or to the decay width.

Algebras introduced in this paper are isomorphic to the standard algebra of complex numbers. This isomorphism has been discussed previously in \cite{Kaniadakis,Scarfone1} for the $\kappa$-case and more in general in \cite{Lobao,Kalogeropoulos}. In the present formalism, the isomorphism is realized according to Equations (\ref{ff2}) and (\ref{f12}), here rewritten as
\begin{eqnarray}
&&x_G+y_G=(x\oplus y)_G\\
&&x^F\cdot y^F=(x\,\tilde\otimes\,y)^F
\end{eqnarray}
On the physical ground, generalized algebras may be related to the interaction present in the system. This idea has been developed in the past by several authors (\cite{Lavagno,Zhang,Quarati1,Wang1,Biro,Biro1,Scarfone10,Lenzi2} to cite a few) and is founded on the possibility to interchange the role of statistical correlations with interactions. In fact, it is known that when a system is weakly interacting, the BG distribution describes well its mechanical statistics proprieties. It turns out that, if interactions can be neglected, \emph{i.e}., the energy of the whole system factorizes in the sum of the energies of the single parts: $U^{{\rm A}+{\rm B}}=U^{\rm A}+U^{\rm B}$, distribution factorizes too: $p^{{\rm A}+{\rm B}}=p^{\rm A}\cdot p^{\rm B}$. This is a consequence of the exponential form of BG distribution being $\exp(x+y)=\exp(x)\,\exp(y)$. On the other hand, if interactions are not negligible $U^{{\rm A}+{\rm B}}=U^{\rm A}+U^{\rm B}+U^{\rm int}$, BG distribution cannot be factorized into the product distribution. Correlations among the parts of the system can occur. A possibility to treat these systems is by introducing a different entropic form whose related distribution is able to capture the salient statistical features, characterized by the underlying   algebraic structure.

Such interpretation can be summarized by means of Equations (\ref{f1}) and (\ref{f11}), which, opportunely relabeled, becomes
\begin{eqnarray}
&&p(\tilde x_i+\tilde y_j)=p(\tilde x_i)\,\tilde\otimes\,p(\tilde y_j)\\
&&p(\tilde x_i\oplus \tilde y_j)=p(\tilde x_i)\,p(\tilde y_j)
\end{eqnarray}
where suitable deformed sum and product can describe approximatively the interactions and the correlations between the parts of the systems.

Finally, let us stress again about the generality of the method proposed in this paper. Algebraic structures arise from the existence of a complex function $g(z)$ and its inverse function univocally defined in the complex plane. Applications to statistical mechanics require that generator $g(z)$ takes real values in a subset of the real axes including the interval $[0,\,1]\in U$, where probabilities make sense. Conditions on $g(x)$, for $x\in U$, can be imposed for physical reasons. In this paper we have discussed about the existence of an entropic form related to these algebras, although other conditions should be imposed on $g(x)$ if necessary. For instance, if one is interested in studying the thermodynamical equilibrium, a concavity condition on $g(x)$ may be imposed. However, these conditions are not pertinent to the construction of the algebraic structures. In fact, the proposed method can be successfully applied to introduce algebras starting from generalized distributions with or without a related entropic form. For instance, algebraic structure of the kind discussed in this paper can be derived for the three parameter distribution introduced in \cite{Kaniadakis5} or the one parameter distribution related to the generalized exponential $\exp^\kappa(x)=\exp({\rm arctan}(\kappa\,x)/\kappa)$ introduced in \cite{Kaniadakis4}.

\app\sect{Proof of Theorem 1}

\noindent For the sum $\oplus$ defined in Equation (\ref{s1}) we have
\begin{eqnarray}
\nonumber
&&{\rm Commutativity}:\\
&&\hspace{5mm}z\oplus z'=f\big(\ln\big(\exp g(z)+\exp g(z')\big)\big)=f\big(\ln\big(\exp g(z')+\exp g(z)\big)\big)=z'\oplus z\hspace{15mm}\\
\nonumber
\nonumber
&&{\rm Associativity}:\\
\nonumber
&&\hspace{5mm}z\oplus(z'\oplus z'')=z\oplus f\big(\ln\big(\exp g(z')+\exp g(z'')\big)\big)\\
\nonumber
&&\hspace{23mm}=f\big(\ln\big(\exp g(z)+\exp g(z')+\exp g(z'')\big)\big)\\
\nonumber
&&\hspace{23mm}=f\big(\ln\big(\exp g(z)+\exp g(z')\big)\big)\oplus z''\\
\nonumber
&&\hspace{23mm}=(z\oplus z')\oplus z''\\
\nonumber
\nonumber
&&{\rm Neutral \ element}:\\
&&\hspace{5mm}z\oplus O=f\big(\ln\big(\exp g(z)+\exp g(O)\big)\big)=f\big(\ln\big(\exp g(z)\big)\big)=z\\
\nonumber
&&\hspace{-8mm}{\rm where \ }O=\lim_{z\to-\infty}f(z),\\
\nonumber
\nonumber
&&{\rm Opposite}:\\
\nonumber
&&\hspace{5mm}z\oplus [-z]=f\big(\ln\big(\exp g(z)+\exp g([-z])\big)\big)\\
\nonumber
&&\hspace{19mm}=f\big(\ln\big(\exp g(z)-\exp g(z)\big)\big)\\
&&\hspace{19mm}=f(-\infty)=O
\end{eqnarray}
where $[-z]=f\big(\ln\big(-\exp g(z)\big)\big)$.

\hspace{5mm}

\noindent For the product $\otimes$ defined in Equation (\ref{p1}) we have
\begin{eqnarray}
\nonumber
&&{\rm Commutativity}:\\
&&\hspace{5mm}z\otimes z'=f\big(g(z)+g(z')\big)=f\big(g(z')+g(z)\big)=z'\otimes z\hspace{55mm}\\
\nonumber
\nonumber
&&{\rm Associativity}:\\
\nonumber
&&\hspace{5mm}z\otimes(z'\otimes z'')=z\otimes f\big(g(z')+g(z'')\big)\\
\nonumber
&&\hspace{23mm}=f\big(g(z)+g(z')+g(z'')\big)\\
\nonumber
&&\hspace{23mm}=f\big(g(z)+g(z')\big)\otimes z''\\
&&\hspace{23mm}=(z\otimes z')\otimes z''
\\
\nonumber
\nonumber
&&{\rm Identity \ element}:\\
&&\hspace{5mm}z\otimes 1=f\big(g(z)+g(1)\big)=f\big(g(z)\big)=z\\
\nonumber
&&
\nonumber {\rm Inverse}:\\
&&\hspace{5mm}
x\otimes[1/z]=f\big(g(z)+g([1/z])\big)=f\big(g(z)-g(z)\big)=f(0)=1
\end{eqnarray}
where $[1/z]=f\big(-g(z)\big)$.

\noindent Finally
\begin{eqnarray}
\nonumber
&&{\rm Distributivity}:\\
\nonumber
&&\hspace{5mm}z\otimes(z'\oplus z'')=z\otimes f\big(\ln\big(\exp g(z')+\exp g(z'')\big)\big)\hspace{50mm}\\
\nonumber
&&\hspace{23mm}=f\big(g(z)+\ln\big(\exp g(z')+\exp g(z'')\big)\big)\\
\nonumber
&&\hspace{23mm}=f\big(\ln\big(\exp g(z))+\ln(\exp g(z')+\exp g(z'')\big)\big)\\
\nonumber
&&\hspace{23mm}=f\big(\ln\big(\exp g(z)\cdot(\exp g(z')+\exp g(z''))\big)\big)\\
\nonumber
&&\hspace{23mm}=f\big(\ln\big(\exp g(z)\cdot\exp g(z')+\exp g(z)\cdot\exp g(z'')\big)\big)\\
\nonumber
&&\hspace{23mm}=f\big(\ln\big(\exp(g(z)+g(z'))+\exp(g(z)+g(z''))\big)\big)\\
\nonumber
&&\hspace{23mm}=f\big(\ln\big(\exp(g(z\otimes z'))+\exp(g(z\otimes z''))\big)\big)\\
&&\hspace{23mm}=(z\otimes z')\oplus(z\otimes z'')
\end{eqnarray}

\sect{Proof of Theorem 2}

\noindent For the sum $\tilde\oplus$ defined in Equation (\ref{s2}) we have
\begin{eqnarray}
\nonumber
&&{\rm Commutativity}:\\
&&\hspace{5mm}z\,\tilde\oplus\, z'=g\big(f(z)\cdot f(z')\big)=g\big(f(z')\cdot f(z)\big)=z'\,\tilde\oplus\, z\hspace{70mm}\\
\nonumber
\nonumber
&&{\rm Associativity}:\\
\nonumber
&&\hspace{5mm}z\,\tilde\oplus\,(z'\,\tilde\oplus\,z'')=z\,\tilde\oplus\, g\big(f(z')\cdot f(z'')\big)=g\big(f(z)\cdot f(z')\cdot f(z'')\big)\\
\nonumber
&&\hspace{23mm}=g\big(f(z)\cdot f(z')\big)\,\tilde\oplus\,z''\\
&&\hspace{23mm}=(z\,\tilde\oplus\,z')\,\tilde\oplus\,z''\\
\nonumber
&&{\rm Neutral \ element}:\\
&&\hspace{5mm}z\,\tilde\oplus\,0=g\big(f(z)\cdot f(0)\big)=g\big(f(z)\big)=z\\
\nonumber
\nonumber
&&{\rm Opposite}:\\
&&\hspace{5mm}z\,\tilde\oplus\,[-z]=g\big(f(z)\cdot f([-z])\big)=g(1)=0
\end{eqnarray}
where $[-z]=g\big(1/f(z)\big)$.

\hspace{5mm}

\noindent For the product $\tilde\otimes$ defined in Equation (\ref{p2}) we have
\begin{eqnarray}
\nonumber
&&{\rm Commutativity}:\\
&&\hspace{5mm}z\,\tilde\otimes\,z'=g\big(\exp\big(\ln f(z)\cdot \ln f(z')\big)\big)=g\big(\exp\big(\ln f(z')\cdot \ln f(z)\big)\big)=z'\,\tilde\otimes\,z\hspace{30mm}\\
\nonumber
\nonumber
&&{\rm Associativity}:\\
\nonumber
&&\hspace{5mm}z\,\tilde\otimes\,(z'\,\tilde\otimes\,z'')=z\,\tilde\otimes\,g\big(\exp\big(\ln f(z')\cdot\ln f(z'')\big)\big)\\
\nonumber
&&\hspace{22mm}=g\big(\exp\big(\ln f(z)\cdot\ln f(z')\cdot\ln f(z'')\big)\big)\\
\nonumber
&&\hspace{22mm}=g\big(\exp\big(\ln f(z)\cdot\ln f(z')\big)\big)\,\tilde\otimes\,z''\\
&&\hspace{22mm}=(z\,\tilde\otimes\,z')\,\tilde\otimes\,z''\hspace{40mm}\\
\nonumber
\nonumber
&&{\rm Identity \ element}:\\
&&\hspace{5mm}z\,\tilde\otimes\,I=g\big(\exp\big(\ln f(z)\cdot\ln f(I)\big)\big)=g\big(f(z)\big)=z
\end{eqnarray}
where $I=g(e)$,
\begin{eqnarray}
\nonumber
&&\hspace{-54mm}{\rm Inverse}:\\
&&\hspace{-49mm}z\,\tilde\otimes\,[1/z]=g\big(\exp\big(\ln f(z)\cdot\ln f([1/z])\big)\big)=g(e)=I \
\end{eqnarray}
where $[1/z]=g(\exp(1/\ln f(z)))$.\\

\noindent Finally,
\begin{eqnarray}
\nonumber
&&\hspace{-32mm}{\rm Distributivity}:\\
\nonumber
&&\hspace{-27mm}z\,\tilde\otimes\,(z'\,\tilde\oplus\,z'')=z\,\tilde\otimes\, g\big(f(z')\cdot f(z'')\big)\\
\nonumber
&&\hspace{-10mm}=g\big(\exp\big(\ln f(z)\cdot\ln(f(z')\cdot f(z''))\big)\big)\\
\nonumber
&&\hspace{-10mm}=g\big(\exp\big(\ln f(z)\cdot\ln f(z')+\ln f(z)\cdot\ln f(z'')\big)\big)\\
\nonumber
&&\hspace{-10mm}=g\big(\exp\big(\ln f(z)\cdot\ln f(z')\big)\cdot\exp\big(\ln f(z)\cdot\ln f(z'')\big)\big)\\
\nonumber
&&\hspace{-10mm}=g\big(f(z\,\tilde\oplus\,z')\cdot f(z\,\tilde\oplus\,z'')\big)\\
&&\hspace{-10mm}
=(z\,\tilde\otimes\,z')\,\tilde\oplus\,(z\,\tilde\otimes\,z'') \
\end{eqnarray}

\sect{Proof of Theorem 3}

\noindent Starting from the functions $g(z)$ and $f(z)$ we define the $g$-numbers
\begin{eqnarray}
z_g=\exp\big(g(z)\big) \
\end{eqnarray}
and the $f$-numbers
\begin{eqnarray}
z^f=\ln\big(f(z)\big) \
\end{eqnarray}
Then, relations (\ref{s1})--(\ref{p1}) and (\ref{s2})--(\ref{p2}) can be rewritten in
\begin{eqnarray}
&&z_g+z'_g=(z\oplus z')_g \ \label{c1}\\
&&z_g\cdot z'_g=(z\otimes z')_g \ \label{c2}
\end{eqnarray}
and
\begin{eqnarray}
&&z^f+z'^f=(z\,\tilde\oplus\,z')^f \ \label{c3}\\
&&z^f\cdot z'^f=(z\,\tilde\otimes\,z')^f \ \label{c4}
\end{eqnarray}
The correspondence between $g$-numbers and $f$-numbers can be established through the solution of~equation
\begin{eqnarray}
z_g=w^f \
\end{eqnarray}
Accounting for Equations (\ref{c1})--(\ref{c4}) we have
\begin{eqnarray}
&&(z\oplus z')_g=(w\,\tilde\oplus\,w')^f \ \\
&&(z\otimes z')_g=(w\,\tilde\otimes\,w')^f \
\end{eqnarray}
for $z_g=w^f$ and $z'_g=w'^f$.

In the case of $f(z)\equiv\exp(z)$ and $g(z)\equiv\ln(z)$ the two algebras collapse each other.


\end{document}